\documentclass[%
 aip,
 amsmath,amssymb,
 reprint,%
]{revtex4-1}

\usepackage{graphicx}
\usepackage{dcolumn}
\usepackage{bm}

\usepackage[utf8]{inputenc}
\usepackage[T1]{fontenc}
\usepackage{mathptmx}
\usepackage{etoolbox}
\usepackage{subcaption}
\usepackage{gensymb}
\usepackage{siunitx}
\usepackage{natbib}

\makeatletter
\def\@email#1#2{%
 \endgroup
 \patchcmd{\titleblock@produce}
  {\frontmatter@RRAPformat}
  {\frontmatter@RRAPformat{\produce@RRAP{*#1\href{mailto:#2}{#2}}}\frontmatter@RRAPformat}
  {}{}
}%
\makeatother
\begin{document}

\preprint{AIP/123-QED}

\title[]{Understanding voltage-controlled magnetic anisotropy effect for the manipulation of dipolar-dominated propagating spin waves}
\author{Adrien. A. D. Petrillo}
\email{a.a.d.petrillo@tue.nl}
 \affiliation{Department of Applied Physics, Eindhoven University of Technology, Eindhoven, The Netherlands}
\author{Mouad Fattouhi}%
\affiliation{ 
Departamento de Física Aplicada, Universidad de Salamanca, 37008 Salamanca, Spain
}%

\author{Adriano Di Pietro}
\affiliation{Istituto Nazionale di Ricerca Metrologica, Strada delle Cacce 91, 10135, Torino, Italy
}%
\affiliation{Politecnico di Torino, Corso Duca degli Abruzzi 24, 10129, Torino, Italy
}%

\author{Marta Alerany Solé}
\affiliation{Department of Applied Physics, Eindhoven University of Technology, Eindhoven, The Netherlands
}%

\author{Luis Lopez Diaz}
\affiliation{ 
Departamento de Física Aplicada, Universidad de Salamanca, 37008 Salamanca, Spain
}%
\author{Gianfranco Durin}
\affiliation{Istituto Nazionale di Ricerca Metrologica, Strada delle Cacce 91, 10135, Torino, Italy
}%

\author{Bert Koopmans}

\affiliation{%
Department of Applied Physics, Eindhoven University of Technology, Eindhoven, The Netherlands
}%

\author{Reinoud Lavrijsen}

\affiliation{%
Department of Applied Physics, Eindhoven University of Technology, Eindhoven, The Netherlands
}%

\date{\today}

\begin{abstract}
Spin waves, known for their ability to propagate without the involvement of moving charges, hold immense promise for on-chip information transfer and processing, offering a path toward post-CMOS computing technologies. This study investigates the potential synergy between propagating Damon-Eshbach spin waves and voltage-controlled magnetization in the pursuit of environmentally sustainable computing solutions. Employing micromagnetic simulations, we assess the feasibility of utilizing spin waves in DE mode in conjunction with localized voltage-induced alterations in surface anisotropy to enable low-energy logic operations. Our findings underscore the critical importance of selecting an optimal excitation frequency and gate width, which significantly influence the efficiency of the phase shift induced in propagating spin waves. Notably, we demonstrate that a realistic phase shift of 2.5$\left[ \pi \ \text{mrad}\right]$ can be achieved at a Co(5nm)/MgO material system via the VCMA effect. Moreover, by tuning the excitation frequency, Co layer thickness, gate width, and the use of a GdO\textsubscript{x} dielectric, we illustrate the potential to enhance the phase shift by a factor of 200 when compared to MgO dielectrics. This research contributes valuable insights towards developing next-generation computing technologies with reduced energy consumption.
\end{abstract}

\maketitle

The use of Magnonics in the context of wave-based computing has gained interest for its potential use as a low Ohmic-loss information carrier. Unlike conventional electronics, where information is conveyed through electronic charges, Magnonics employ spin-waves (SWs) to carry information through the amplitude and phase of a pseudo-particle known as a magnon. This innovative approach may open the road to beyond-CMOS computing technology \cite{Chumak2015-hu,Kruglyak_2010,Chumak_2017} free of Joule heating. As data processing requires phase and amplitude modulation, novel ways of manipulating propagating SWs through a local change of the effective magnetic field have been explored, such as the application of a magnetic field \cite{PhysRevLett.93.047201,Chumak2010-hz,Vogt2014-in}, local laser-induced heating \cite{Vogel2015-xn,Vogel2018-pk,Albisetti2020-tt} or current \cite{PhysRevLett.107.146602,Liu2019-tq,Sarker2023-bv}. In particular, control of phase or amplitude by Oersted field or SW current has been a topic of study for the development of Mach-Zehnder spin-wave interferometer architecture\cite{10.1063/1.2834714,10.1063/1.2975235,Khitun_2010,Khitun_2010}. In an effort to further decrease the energy consumption of magnetic materials-based computing devices, voltage-controlled magnetic anisotropy (VCMA) holds great promise, potentially reducing the energy consumption of storage and logic devices by a factor of 100 \cite{Matsukura2015-bo}. Effects of an electric field on surface anisotropy (SA) have widely been studied over the recent years \cite{doi:10.1126/science.1136629,10.1063/1.3429592,PhysRevLett.113.267202,10.1063/1.4961621,Shiota2012-mv,Chiba2011-im,Wang2012-pw,Maruyama2009-dj,Bauer2015-dv} and, in essence, can be used to manipulate the spin-wave (magnon) dispersion relation via two routes: a fast, volatile, electronic effect \textit{e.g.}, direct solid-state gating, or a slow, non-volatile, magneto-ionic effect \textit{e.g.}, moving mobile ions in or out of the ferromagnetic layers. The magneto-ionic effect is a process of electrochemical nature happening at ferromagnetic/oxide (FM/O) interfaces\cite{RevModPhys.89.025008}, where oxygen ions are brought to the interface between the oxide dielectric and the ferromagnetic layer. Due to coupling between 2p orbitals of oxygen ions and 3d orbitals from Co at the interface\cite{PhysRevB.84.054401}, changes in charge densities lead to a change of the magnetic anisotropy via a modulated spin-orbit coupling. The magneto-electronic effect has been interpreted as a change of the electronic occupation state at the interface between the dielectric and the ferromagnetic layer \cite{PhysRevLett.101.137201,PhysRevLett.102.187201,PhysRevLett.102.247203} and electric-field-induced dipole \cite{Miwa2017-az}. While volatile, the latter effect is particularly suited for logic due to the fast transport of the charges at the interface, compared to magneto-ionic effects befitting for storage devices due to their non-volatility but slower response due to the electrochemical nature. 

\indent Theoretical and experimental studies based on VCMA control of SWs include reconfigurable magnonic crystals\cite{PhysRevB.95.134433}, nanochannels\cite{PhysRevApplied.9.014033,PhysRevB.95.134433} and SW phase shifters \cite{PhysRevB.98.024427,10.1063/1.5037958}. Moreover, these studies have primarily been focusing on insulating ferromagnetic SW conduits of yttrium iron garnet (YIG)\cite{PhysRevLett.113.037202,10.1063/1.2198111}. In particular, studies have concentrated on the Damon-Eshbach (DE) SW mode configuration. Indeed, due to the localization of DE SWs at the interfaces of the ferromagnetic layer in direct contact with the dielectric where the VCMA effect is induced, the VCMA-induced non-reciprocal transport of the SWs can efficiently be probed using a phase-sensitive all-electrical method\cite{Kasukawa_2018,Nawaoka_2015,10.1063/1.4914060}. In these studies, the VCMA effect was assumed to be applied homogeneously along the entire spin-wave conduit. Hence, the question of what happens when the VCMA effect is applied locally along the path of a spin-wave conduit remains open. 

\indent In this study, we present an investigation of the impact of locally applied electric fields at the interface between a Co ferromagnetic layer and a MgO dielectric on dipole-dominated DE propagating SWs within thin-film SW-conduits, employing Mumax\textsuperscript{3} micromagnetic simulations\cite{10.1063/1.4899186}. Initially, we elucidate the influence of SA modifications on the dispersion characteristics of the excited modes, shedding light on their implications for the wavevector of SWs both entering and exiting the VCMA-gated region. Subsequently, we examine the propagation characteristics of the resulting scattered SWs. Furthermore, we analyze the frequency and gate width-dependent VCMA-induced phase shifts in the scattered SWs at a local scale. Finally, we discuss the feasibility of employing the DE configuration for SW phase shifting with a metallic Cobalt ferromagnetic waveguide and optimizing the ferromagnetic/dielectric interface.

\begin{figure}[h]
    \centering    
           \includegraphics[width=1.0\linewidth]{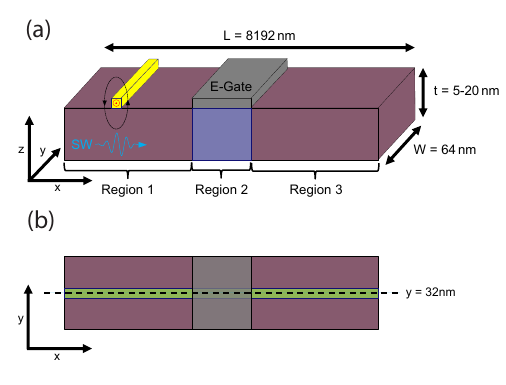}       
     \caption{(a) Schematic of the system under study. The strip is divided into three regions. The SW is excited in region 1, $K_{S}$ is modified within region 2, under the gate, and the SW is detected in region 3. (b) Top view of the simulation box. A row of cells (in green) of $2\times2\times2.5$ nm centered around the center of the strip (\textit{y}=32nm) along which the magnetization is sampled.}
    \label{fig:stack}
\end{figure}

Our model system is an 8192 nm long (\textit{x}) and 64 nm (\textit{y}) wide Co waveguide with a thickness of 5, 10, and 20 nm (\textit{z}). The simulation box is subdivided in 2 $\times$ 2 $\times$ 2.5 nm cells in the \textit{x}-, \textit{y}-, \textit{z}- directions. Due to the narrowness of our waveguide and our focus on the lowest frequency DE mode i.e, the wavevector along (\textit{y}) $k_{y}$=0, Periodic Boundary Conditions (PBCs) are applied along \textit{y}. The used magnetic parameters for Cobalt are $\gamma$ = $1.9\times10^{11}$ Hz rad/T, $M_{s}$ = $1.42\times10^{6}$ A/m, $A_{ex}$ = $20\times10^{-12}$ J/m, and $\alpha$ = 0.005 respectively, representing the gyro-magnetic ratio, the saturation magnetization, the exchange stiffness and the damping, respectivelly. No uniaxial anisotropy is introduced. Exponentially increasing spatial damping profiles over a distance of 1.016 $\mu \textrm{m}$ are introduced at both ends (\textit{x}) of the simulation box to prevent SW reflections. DE SWs are excited by applying a static transverse magnetic field of 100 mT along \textit{y}, fully saturating the magnetization in the \textit{y}-direction and applying a sinusoidal AC magnetic field with a Gaussian profile varying in the \textit{x} direction for a time period of $t=10$ ns, leading to a spin-wave with small time-dependent dynamic magnetization components in the \textit{x-z} plane. 

In the following, we define three regions (See Fig. \ref{fig:stack}a). In region 1 and 3, the base anisotropy $K_{S_{E=0}} = 0.71\times10^{-3}$J/m\textsuperscript{2} is chosen following Kasukawa \textit{et. al.}\cite{Kasukawa_2018}, with $E=0$ indicating the nominal anisotropy of the waveguide, i.e., when no VCMA is applied. In region 2, we emulate the effect of VCMA, \textit{e.g.}, by solid-state gating, by modifying the magnetic anisotropy to $K_{S_{E \neq 0}}$. In region 1, an AC field located 1.2 $\mu \textrm{m}$ away from the left edge of the waveguide is responsible for exciting SWs propagating to the left (-\textit{x}) and right (+\textit{x}) directions. The SWs traveling to the right first encounter region 2 and finally propagate to region 3, where we detect the SWs after scattering twice at the entry and exit of region 2 due to the modified SA $K_{S_{E \neq 0}}$ in that region. The SW propagating to the left of the waveguide is absorbed due to the exponentially increasing $\alpha$, which prevents SW reflection at the left edge. 

\begin{figure}[h!]
    \centering    
           \includegraphics[width=1.0\linewidth]{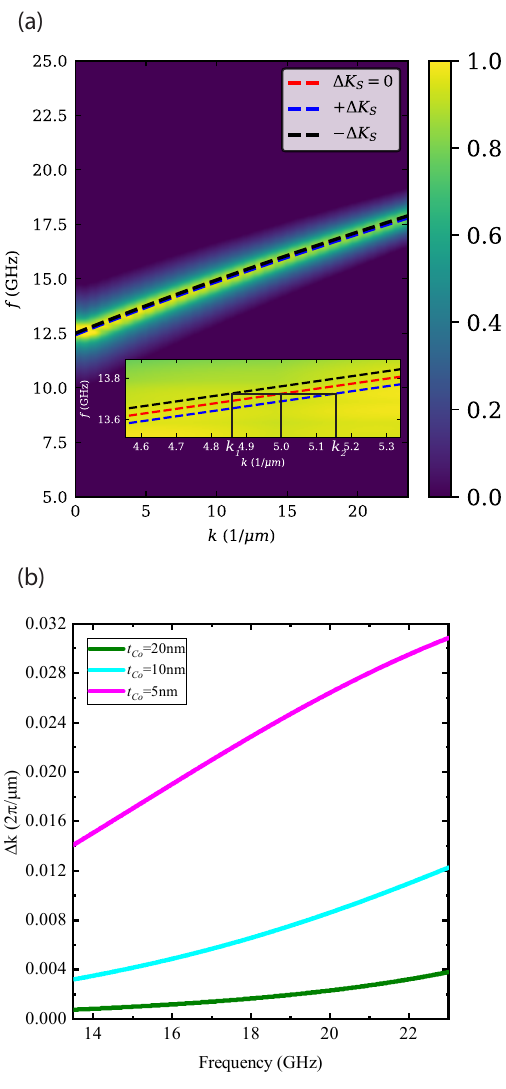}       
     \caption{(a) FFT of M\textsubscript{x}/M\textsubscript{S} component of Micromagnetic simulation for a Co($t_{Co}=5$ nm) waveguide. The inset shows the effect of a change of $K_{S}$ on the dispersion from equation \ref{eq1}. (b) Numerically deduced change of wavevector due to the VCMA as a function of frequency for $t_{Co}=$5, 10, and 20 nm waveguides using equation \ref{eq1}.}
    \label{fig:dispersion}
\end{figure}

In this article, data are extracted by defining a row of cells of $2\times2\times2.5$ nm at the top surface of the strip along the x direction of the strip, centered around y = 32 nm (see Figure \ref{fig:stack}b). As the spin-wave profile is uniform across the thickness, we choose this region to minimize simulation time, saving only data from the selected region. The region is selected around the center of the strip to avoid peculiarities in the demagnetizing field due to edge effects. A Fast Fourier Transform (FFT) in space and time of the dynamic magnetization $m_x$ component in this box allows us to obtain the dispersion relation of our system as shown in Figure \ref{fig:dispersion}a. The highest intensity of the FFT follows very closely the analytical dispersion (shown by the red dashed line) as given by \cite{doi:10.1126/sciadv.aba5457} 
\begin{widetext}
    \begin{equation}
    \label{eq1}
         f = \frac{\mu_{0}\gamma}{2\pi}\sqrt{\Big( H+\frac{2A_{ex}}{\mu_{0}M_{S}}k^{2}\Big)\Big(H+\frac{2A_{ex}}{\mu_{0}M_{S}}k^{2}+M_{S}-H_{p}(V_{gate})\Big)+\frac{M_{S}}{4}(1-e^{-2t\lvert k \rvert})(M_{S}- H_{p}(V_{gate}))} ,
    \end{equation}
\end{widetext}
with $H$ the applied static magnetic field, $k$ the SW wavevector and $H_{p}(V_{gate})$ the voltage ($V_{gate}$) dependent out-of-plane SA field. From this excellent correspondence, we conclude that our Micromagnetic simulation setup correctly describes DE-mode SWs in Co-based thin films. Once the SWs are excited in region 1, they propagate to region 2, at the center of the strip, where the effective $K_{S_{E\neq0}} $ is modified to reproduce the VCMA effect induced at the interface between the Co and the MgO. Here, we include the VCMA effect as a linear dependency on the applied electric field \cite{SONG201733}. The voltage-dependent out-of-plane SA $H_{p}(V_{gate})$ can be expressed as $\beta\frac{2K_{S}}{\mu_{0}M_{S}t_{Co}}$ where $\beta$ is a linear term with $V_{gate}$ describing the effect of the electric field on the change of $K_{S}$ and $t_{Co}$ the thickness of the Co layer. Here, we have used the value based on the work of Kasukawa \textit{et al.} multiplied by ten to boost the effect in our simulations, which corresponds to a relative change of the total perpendicular magnetic anisotropy of 0.44\%. Once the SWs leave region 2, they are detected in region 3, where the anisotropy is set to $K_{S_{E=0}}$ as in region 1. In our following simulations, we vary the gate width ($W_{gate}$) from 0 to 2.5 $\mu \textrm{m}$, and $K_{S}$ is increased (resp. decreased) by a maximum amount of 3 mJ/m\textsuperscript{2}.

The effect of the change of $K_{S}$ can be understood from the inset of Figure \ref{fig:dispersion}a, which shows a change in the dispersion, under the gated region, for different $K_{S}$. Once $K_{S}$ is increased by +$\Delta K_S$ (resp. decreased by -$\Delta K_S$), the dispersion relation is shifted down (resp. up). Let us now assume a SW excited with a wavevector $k$ = 5 $\left[2\pi\ \textrm{rad}/\mu \textrm{m}\right]$. An increase of anisotropy +$\Delta K_S$ (resp. decrease of anisotropy -$\Delta K_S$) will result in a scattering of the SW to a wavevector $k_2$ (resp. $k_1$). By numerically solving equation \ref{eq1} for the wavevector $k$, we plot, in fig. \ref{fig:dispersion}b, the change of wavevector $\Delta k =\lvert k_{0\%}-k_{-0.44\%}\rvert$ with $ k_{0\%}$ the initial SA and $k_{-0.44\%}$ the SA corresponding to a decrease of 0.44\% of $ k_{0\%}$. As $K_{S}$ is an interfacial effect, the VCMA effect scales as $1/t_{Co}$. The dependency of $\Delta k$ with increasing frequency can be understood via its dispersion relation. For all thicknesses, $\Delta k$ increases with frequency, which is a direct consequence of the flattening of the dispersion relation at higher frequencies (\textit{e.g.}, Fig \ref{fig:dispersion}a), which increases the change in $\Delta k$ for a given change in $K_S$. The consequence of a change of wavevector is a phase shift of the propagating SW induced in the gated region. Therefore, an increase (decrease) of SA results in a decrease (increase) of wavevector, inducing a positive phase shift due to its higher (lower) group velocity (v\textsubscript{g}). Hence, we conclude that, in accordance with the analytical dispersion, the total amount of acquired phase shift per unit length (due to the $\Delta k$) of a SW traveling through the gated region 2 strongly depends on the conduit layer thickness via its dispersion.

\begin{figure}[h!]
    \centering    
           \includegraphics[width=1.0\linewidth]{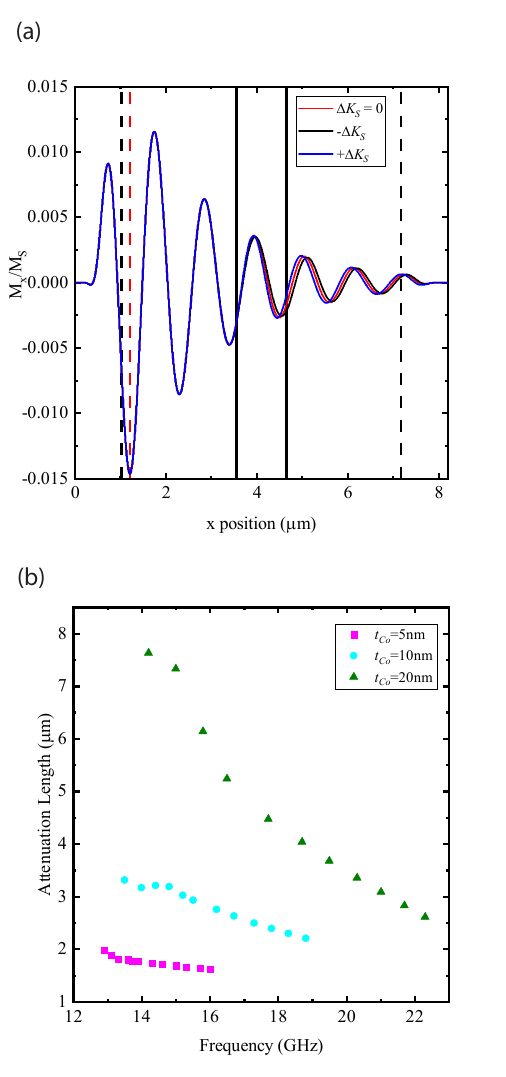}       
     \caption{(a) M\textsubscript{x}/M\textsubscript{S} component of the magnetization along \textit{x} at time instant $t=10$ ns for a $t_{Co}=$5 nm layer for no change of anisotropy in red, a positive change in blue and a negative change in black. Numbers 1, 2, and 3 correspond, respectively, to regions 1, 2, and 3 shown in Fig. \ref{fig:stack} with the region 2 delimited by the 2 continuous black lines. (b) Fitted attenuation length $l_{att}$ ($t=10$ ns) as a function of frequency excitation of SWs for $t_{Co}=$5, 10, and 20 nm.}
    \label{fig:profile}
\end{figure}

Figure \ref{fig:profile}a shows the normalized M\textsubscript{x}/M\textsubscript{S} component of the magnetization along the \textit{x}-axis of the waveguide. To compare our results, we choose the same time instant of the M\textsubscript{x}/M\textsubscript{S} component, corresponding to the last time frame of the simulation ($t=10$ ns). This ensures that the amplitude of M\textsubscript{x}/M\textsubscript{S} over time does not vary and corresponds to the magnetization equilibrium. As described previously, an exponential damping profile is present at both ends of the waveguide, delimited by the black dashed lines (see Fig.\ref{fig:profile}a). As seen in Figure \ref{fig:profile}a, the SW is excited 1.2 $\mu \textrm{m}$ from the left side of the strip, noted by a red dashed line. The two continuous black lines centered around the strip indicate the gated region with a width of $W_{gate} =$ 1.1 $\mu \textrm{m}$. The amplitude of M\textsubscript{x}/M\textsubscript{S} shows the typical exponential decay with distance due to damping. Once $K_{S}$ is decreased (resp. increased) by a factor -$\Delta K_{S}$ (resp. +$\Delta K_{S}$) in the gated region as shown by the blue curves, a positive (resp. negative) phase shift is induced in the gated region, which, after traveling into region 2, results in a constant phase shifted propagating SW that can be measured in region 3. Due to the small change in $k$, no significant back-scatter effects are observed at the interfaces between regions 1 and 2 and region 2 and 3. This simple representation captures, in essence, the effect of VCMA on the phase change of the SWs. On the other hand, using these profiles, we can extract the effect of gating on the SW attenuation length $l_{att}$ as a function of the frequency for different $t_{Co}=$, as shown in Fig. \ref{fig:profile}b. To do so, we fit the M\textsubscript{x}/M\textsubscript{S} component of the magnetization along the \textit{x} with the expression
\begin{equation}
    \label{eq2}
         A\ e^{- l_{att} x}\ \sin(k x+ \phi) ,
\end{equation}
with $A$ the amplitude, $l_{att}$ the attenuation length, $x$ the position along the \textit{x}-axis of the waveguide, $k$ the wavevector and $\phi$ the phase of the SW. As expected, $l_{att}$ decreases with increasing frequency as the dispersion relation flattens off, resulting in a lower v\textsubscript{g}. Moreover, the thinner the Co layer, the lower v\textsubscript{g} in the studied frequency regime, resulting in a smaller attenuation length. Hence, as the SW phase accumulates with the distance traveled, one must carefully balance the attenuation length and accumulated phase shift per unit length.

\begin{figure}[h!]
    \centering    
           \includegraphics[width=1.0\linewidth]{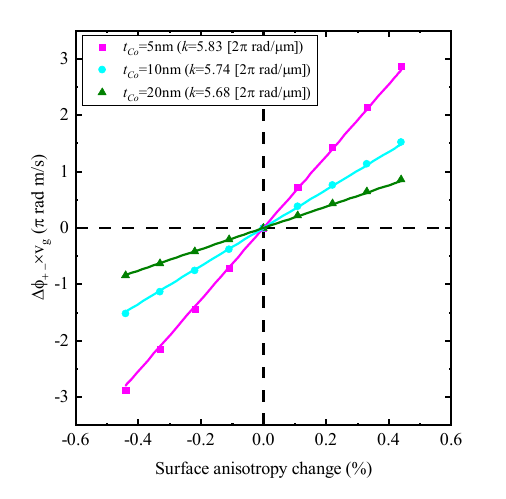}       
     \caption{Phase shift ($t=10$ ns) multiplied by the frequency-dependent v\textsubscript{g} as a function of $K_{S}$ change for $t_{Co}=$ 5, 10, and 20 nm from Micromagnetic simulations (symbols) and analytical model (continuous lines).}
    \label{fig:5}
\end{figure}

In order to extract the accumulated phase of the propagating SW in region 3, resulting from the local VCMA-induced change of wavevector in region 2, we use equation \ref{eq2} to fit the portion of the SW in region 3 of interest. Here we define the phase shift as $\Delta \phi_{+-} = \lvert \phi_{0\%}-\phi_{+-0.44\%} \rvert$ i.e., the difference between the phase of the SW with wavevectors $k_{0\%}$ and $k_{+0.44\%}$ and $k_{0\%}$ and $k_{-0.44\%}$. To compensate for the thickness-dependent change in accumulated phase due to the change in v\textsubscript{g}, in Fig. \ref{fig:5} we plot, $\Delta k$ multiplied by the wavevector-dependent v\textsubscript{g} as extracted from the dispersion relation. As expected from the model from equation \ref{eq1}, the VCMA-induced phase shift scales linearly with the change of $K_{S}$. Moreover, we identify the necessity to multiply the phase by the v\textsubscript{g} in order for the accumulated phase to scale as 1/$t_{Co}$. This suggests an effect of VCMA on the phase inversely proportional to v\textsubscript{g}, implying a time-dependent effect of VCMA on the propagating SW, in the gated region.


\begin{figure}[h!]
    \centering    
           \includegraphics[width=1.0\linewidth]{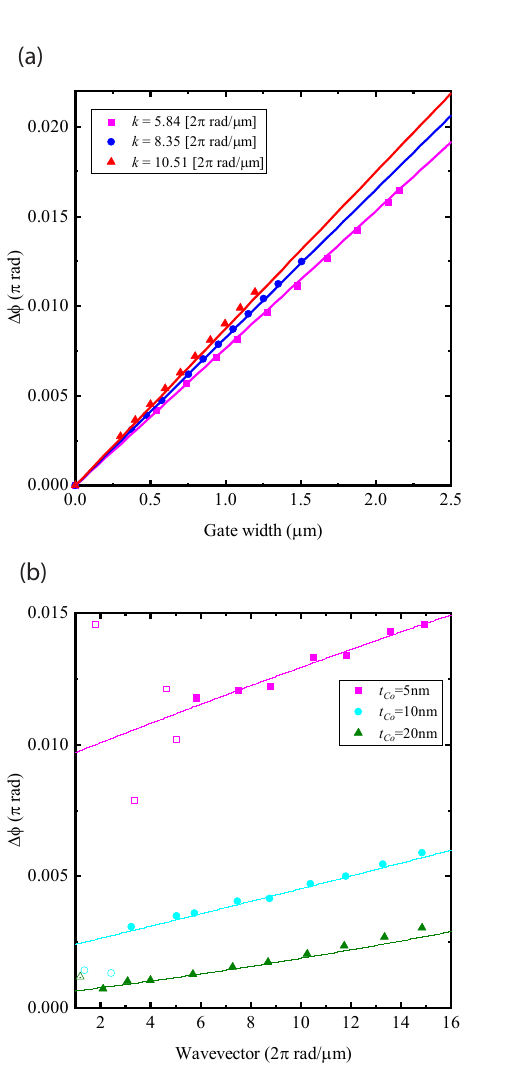}       
     \caption{(a) Phase shift ($t=10$ ns) as a function of the $W_{gate}$ for a $t_{Co}=$ 5 nm waveguide and wavevectors 5.84, 8.35, and 10.51 $\left[2\pi\ \textrm{rad}/\mu \textrm{m}\right]$ as obtained from Micromagnetic simulations (symbols) and analytical model (solid lines). (b) Phase shift ($t=10$ ns) as a function of the SW wavevector for a $t_{Co}=$ 5 nm waveguide from Micromagnetic simulations (symbols) and analytical model (solid lines). A discrepancy between the Micromagnetics and analytical model is observed for lower wavevector as denoted by the open symbols. }
    \label{fig:figure5}
\end{figure}

Performing full logic operations with SWs requires a $\pi$ rad shift of the SW compared to a given SW reference signal. Although $t_{Co}=$ can be reduced to boost the VCMA effect and, therefore, the phase shift, a reduction in thickness is limited in its efficiency due to the decreased $l_{att}$. Fig. \ref{fig:figure5}a shows the phase shift $\Delta \phi = \lvert \phi_{0\%}-\phi_{-0.44\%}\rvert$ as a function of the gate length for wavevectors of 5.84, 8.35, and 10.51 $\left[2\pi\ \textrm{rad}/\mu \textrm{m}\right]$ for a Co layer of $t_{Co}=5$ nm. As expected from the analytical model from equation \ref{eq1}, we observe a linear increase of the $\Delta \phi$ with increasing $W_{gate}$. Good correspondence between the analytical model ($\Delta k \times W_{gate}$) and the simulations show that the analytical model can accurately predict phase changes induced by VCMA. 

To gain further insight into the frequency dependency of the VCMA effect, we plot the $\Delta \phi$ as a function of the SW frequency in Fig. \ref{fig:figure5}b. Indeed, as v\textsubscript{g} decreases with SW frequency and Co layer thickness, the accumulated phase increases for a given $W_{gate}$. A discrepancy between the Micromagnetic and analytical model is observed for lower wavevector (open symbols) and is currently speculatively attributed to reflections and/or finite size effect of the simulation box which becomes of the order of the SW wavelength. The linear trend in the VCMA-dependent $\Delta \phi$ with the excitation frequency and $W_{gate}$ allows for a discussion of the feasibility of a $\pi$ rad phase shift induced by a local changes of $K_{S}$ for the application of logic operation. Let us assume, as an example, a wavevector k = 8.35 $\left[2\pi\ \textrm{rad}/\mu \textrm{m}\right]$ and $W_{gate}$ of 1.5 $\mu \textrm{m}$. Assuming minimum reflections from the gated area and no resonance in the gated region due to the SW wavelength being commensurable with the gated region width, we can naively deduce, from Figure \ref{fig:figure5}a, that the maximum possible achievable $\Delta \phi$, in these conditions, is of the order of 12.6 $\left[ \pi \ \text{mrad}\right]$ when using a MgO dielectric as a catalyst for the change of $K_{S}$. As the changes of SA, in this work, were assumed to be Kasukawa \textit{et.al.}. multiplied by ten to boost the effect in our simulations, a $\Delta \phi$ of 1.26 $\pi$ mrad can be expected for a realistic, non-boosted, VCMA-induced $K_{S}$ change. It is important to note that as a positive change of $K_{S}$ will induce an additional $\Delta \phi$ of 1.26 $\left[ \pi \ \text{mrad}\right]$ due to the symmetrical $\Delta \phi$ induced by a change of SA, the total $\Delta \phi$, in these conditions, would be 2.5 $\left[ \pi \ \text{mrad}\right]$. While we think this phase change is commensurable with the limit of phase sensitivity of all-electrical propagating SW spectroscopy measurement methods \cite{10.1063/1.1597745}, an optimization of the excitation frequency and $W_{gate}$ is needed to achieve phase changes of the order of $\pi$ rad phase shift. Although MgO does not allow for significant changes of $K_{S}$, optimizing the dielectric layer may considerably increase the efficiency of VCMA-induced $\Delta \phi$ with DE SWs. One possibility is the use of a GdO\textsubscript{x} dielectric, which showed an efficiency of change of $K_{S}$ of the order of 500 $fJ/(V\cdot \textrm{m})$ at Co($t_{Co}$=0.9nm)/GdO\textsubscript{x}($t_{GdO}=$3nm) interfaces \cite{Bauer2015-pf}. This efficiency in VCMA is 200 times superior to MgO, which could potentially increase the $\Delta \phi$ to a value of 0.25 $\left[ \pi \ \text{rad}\right]$. By tuning the applied voltage within the breakdown limit of GdO\textsubscript{x} and optimizing the excitation frequency and $W_{gate}$, we could, therefore, imagine further improvement in the efficiency of the $\Delta \phi$. However, the use of GdO\textsubscript{x} presents a significant limitation in time efficiency due to the oxidation mechanism responsible for the change of $K_{S}$. On top of tuning the parameters previously described, the efficiency of VCMA of $\Delta \phi$ is also shown to scale as $1/t_{Co}$, meaning that a thinner layer would improve further the efficiency of VCMA and get closer to a $\pi$ rad phase shift.  It is important to note that the effect of VCMA on interfacial Dzyaloshinskii-Moriya interaction (iDMI) has not been included in this study, but further research in the combination of VCMA effect on $K_{S}$ and iDMI may lead to even more significant changes of phase as VCMA effect have been reported on iDMI \cite{Nawaoka_2015}.

In summary, we have investigated the effect of a local change of $K_{S}$ on the phase of propagating SWs in the DE configuration. We have shown a linear dependency of the VCMA-induced $\Delta \phi$ with the excitation frequency and $W_{gate}$ in accordance with analytical models. Moreover, we showed the importance of the v\textsubscript{g} in the expected $\Delta \phi$ when studying the effect of a change of $t_{Co}$. We determined that the maximum realistic $\Delta \phi$ in Co($t_{Co}$=5nm)/MgO stack is of the order of 2.5 $\left[ \pi \ \text{mrad}\right]$ and can be further improved by increasing the frequency excitation, $W_{gate}$, decreasing $t_{Co}$ and by opting for a GdO\textsubscript{x} dielectric, the latter allowing to improve the efficiency of change of phase by a factor 200 compared to what can be expected from a MgO dielectric.

\begin{acknowledgments}
This project has received funding from the European Union’s Horizon 2020 research and innovation programme under the Marie Skłodowska-Curie grant agreement No. 860060 “Magnetism and the effect of Electric Field” (MagnEFi). 
\end{acknowledgments}

\section*{Data Availability Statement}
The data that support the findings of this study are available from the corresponding author upon reasonable request.

\nocite{*}
\bibliography{aipsamp}
\end{document}